\documentclass[showpacs,aps,pra,amssymb,floatfix]{revtex4-1}

\usepackage{graphicx}
\usepackage{setspace}
\usepackage{epsfig}
\usepackage{amsmath}
\usepackage{color}
\usepackage{xcolor}
\usepackage{subfig}
\usepackage{float}
\usepackage[normalem]{ulem}
\usepackage{hyperref}
\hypersetup{
  colorlinks   = true, 
  urlcolor     = black, 
  linkcolor    = red, 
  citecolor    = blue 
}
\definecolor{verde}{rgb}{0.04, 0.6, 0.02}
\definecolor{violeta}{rgb}{0.6, 0, 0.6}
\definecolor{cinza}{rgb}{0.6, 0.6, 0.6}

\begin{document}

\title{Lyapunov stability under $q$-dilatation and $q$-contraction of coordinates}
\author{Tulio M Oliveira$^1$, Vinicius Wiggers$^1$, Eduardo Scafi$^1$,  Silvio Zanin$^1$, Cesar Manchein$^2$ and Marcus Werner Beims$^1$}
\email{mbeims@fisica.ufpr.br}
\address{$^1$Departamento de F\'{\i}sica, Universidade Federal do
Paran\'a, 81531-980 Curitiba, PR, Brazil}
\address{$^2$Departamento de F\'{\i}sica, Universidade do Estado de Santa Catarina,  89219-710 Joinville, SC, Brazil}

\begin{abstract}
This study examines the Lyapunov stability under coordinate $q$-contraction and $q$-dilatation in three dynamical systems: the discrete-time dissipative Hénon map, and the conservative, non-integrable, continuous-time Hénon-Heiles and diamagnetic Kepler problems. The stability analysis uses the $q$-deformed Jacobian and $q$-derivative, with trajectory stability assessed for $q > 1$ (dilatation) and $q < 1$ (contraction). Analytical curves in the parameter space mark boundaries of distinct low-periodic motions in the Hénon map. Numerical simulations compute the maximal Lyapunov exponent across the parameter space, in Poincaré surfaces of section, and as a function of total energy in the conservative systems. Simulations show that $q$-contraction ($q$-dilatation) generally decreases (increases) positive Lyapunov exponents relative to the $q = 1$ case, while both transformations tend to increase Lyapunov exponents for regular orbits. Some exceptions to this trend remain unexplained regarding Kolmogorov-Arnold-Moser (KAM) tori stability.

\end{abstract}

\maketitle

\section{INTRODUCTION}

Several important functions used in physics and applications can be expressed in terms of hypergeometric series $F(a,b,c,z)$, where  $a,b,c$ are parameters and $z$ ithe variable. These series were presented by Gauss in 1812 and are frequently called Gauss series \cite{gauss1813}. Thirty-four years after Gauss' presentation, Heine introduced the $q$-hypergeometric series $\phi(a,b,c,q,z)$ \cite{heine46,heine47,heine78}, with the additional parameter $|q|<1$, which tend to the Gauss' series when $q\to1$ since $\lim_{q\to1} (1-q^a)/(1-q)=a$. The $\phi(a,b,c,q,z)$ are $q$-deformed series, which lead to many $q$-deformed functions, which have been used in many research fields.  $F(a,b,c,z)$ and $\phi(a,b,c,q,z)$ are convergent series for $|z|<1$ and $|z|=1$. The connection between the Gauss series and the $q$-hypergeometric series, together with a complete description of the latter, is presented in Ref.~\cite{gasper97}. 

For decades, researchers have searched for a clear physical interpretation of the parameter $q$, ranging from mathematical physics, statistical physics, and quantum physics to relativistic mechanics. For example, the $q$-deformed exponential function was used to show the power-law sensitivity to the initial conditions at the edge of chaotic motion \cite{tsallis97,lyra97}. A crucial step was to recognize the relation between $q$-deformation and the fractal properties of functions \cite{ayse97} and its relation to the fractal dimension of a dissipative attractor \cite{lyra98}. Later, the $q$-deformed quadratic map \cite{sinha05} and also the $q$-deformed Gaussian map \cite{sud09} have been considered.  According to \cite{ayse97}, $q$ is related to the \textit{dilatation} ($q>1$) or the \textit{contraction} ($q<1$) of coordinates and, therefore, expected to be associated with fractal dimensions. In this context, the $q$-deformed Jacobian of the one-dimensional quadratic map was proposed to analyze the stability of periodic orbits under dilatation and contraction of the coordinate \cite{rech10}. The present contribution is an extension of \cite{rech10} and considers the Lyapunov stability under contraction and dilatation of coordinates in the paradigmatic two-dimensional dissipative Hénon map and in the more realistic physical systems, namely the two-dimensional Hénon-Heiles and the diamagnetic Kepler problem. Such $q$-Lyapunov stability analysis is performed by constructing the $q$-deformed Jacobian of each system considering the $q$-derivative presented in \cite{ayse97}, defined as:
\begin{equation}
    \partial_{x_i}^{(q)} f(\textbf{x})\equiv \frac{f(x_1;\ldots;qx_i;\ldots x_N)-f(x_1;\ldots; x_i;\ldots x_N)}{(q-1)x_i},
    \label{df}
\end{equation}
where $\textbf{x} = (x_1;\ldots; x_N)$ is a $N$-dimensional vector.
 The derivative (\ref{df}) measures the rate of changes of the function $f(x_1;\ldots; x_i; ...;x_N)$ concerning a dilatation (or contraction) of the variable $x_i$ by a factor of $q$. In the context of the linear stability analysis in dynamical systems, Eq.~(\ref{df}) can be used to study the stability of periodic points under a dilatation (or contraction) of the dynamical variable $x_i$. In the limit $q\to 1$, the usual derivative is recovered, and therefore, the usual stability under translation of variables. For the implementation of the derivative (\ref{df}) we use \cite{gasper97,ayse97}
\begin{equation}
    \partial_{x_i}^{(q)} f(x_1;\ldots; x_N) = [\psi_i]_q\frac{f(x_1;\ldots; x_N) }{x_i},
    \label{dfPsi}
\end{equation}
where
\begin{equation}
    [\psi_i]_q\equiv \frac{q^{\psi_i}-1}{q-1},
\end{equation}
and $\psi_i$ is the degree of the, \textit{in the coordinate} $x_i$, homogeneous function $f(x_1;\ldots; x_N)$.

For applications, we extend the above definition for functions with distinct power degrees in $x_i$ by using
\begin{equation}
    \partial_{x_i}^{(q)} f(x_1;\ldots; x_N) = \sum_j^g\,[\psi^{(j)}_i]_q\frac{f(x_1;\ldots; x_N) }{x_i},
    \label{dfPsi}
\end{equation}
where the sum is over the $g$ distinct powers of the variable $x_i$ which appear inside $f(x_1;\ldots; x_N) $ and,
\begin{equation}
    [\psi^{(j)}_i]_q\equiv \frac{q^{\psi^{(j)}_i}-1}{q-1}.
\end{equation}
For example, for $f(x,y) = x^2y + x^4y^3$ (using $x_1=x, x_2=y$), we have $j=2,4$ so that  $\partial_{x}^{(q)}f(x,y) = [\psi^{(2)}_x]_q\,xy+[\psi^{(4)}_x]_q\,x^3y^3$. In the limit $\lim_{q\to1}[\psi^{(2)}_x]_q=2$ and  $\lim_{q\to1}[\psi^{(4)}_x]_q=4$ we obtain the usual derivative, as expected. Worthy of mentioning is that the $q$-derivative only provides new effects when nonlinear terms are present since for linear terms, we have $[\psi^{(1)}_x]_q=1$, independent of $q$.

The presentation of this work is divided as follows: Section \ref{Jac} discussed the general procedure to obtain the Lyapunov stability under dilatation and contraction of coordinates. Sections \ref{HM}, \ref{HH}, and \ref{DD} discuss the analytical and numerical results of the dilatation/contraction Lyapunov stability for the Hénon map, Hénon-Heiles potential and the diamagnetic problem, respectively. Section \ref{conclusion} summarizes our main results.

\section{$q$-Lyapunov stability}
\label{Jac}

The usual way to study the Lyapunov stability of trajectories in a dynamical system is to apply a small perturbation to a main trajectory and compare the evolution of the perturbed and non-perturbed trajectories. Roughly saying, the exponential divergence $\sim e^{\lambda t}$ (if it exists) between both trajectories furnishes the Lyapunov exponent $\lambda$ (LE). Formally, consider a $N$-dimensional nonlinear dynamical system described by the differential equations
\begin{eqnarray}
    \frac{d{\bf x}}{dt} = f({\bf x}).
    \label{xdot}
\end{eqnarray}
To study the stability of a solution ${\bf x}(t)$ under a small perturbation (${\bf y}$) of initial conditions, we consider the linear system of differential equations  
\begin{equation}
    \frac{d{\bf y}}{dt} = J(t)\,{\bf y},
    \label{ydot}
\end{equation}  
where the Jacobian is given by $J(t)=F'({\bf x}(t))$, and the prime means the partial derivative relative to the coordinates.  Solving for stability under small perturbations in the linear system (\ref{ydot})  can indeed help understand stability in the nonlinear system (\ref{xdot}) as well. This reduction can simplify the analysis and provide insights into the overall system behavior. All the information about the linear stability of trajectories is given by the eigenvalues of the Jacobian matrix. This analysis is the subject of study in the Lyapunov stability theory \cite{pesin}, and the LE is obtained from the eigenvalues of the Jacobian matrix. Similar arguments can be used to construct the Jacobian ($J_n$) for dynamical systems described by maps ${\bf x}_{n+1}=f({\bf x}_n)$, where $n=0,1,2,\ldots$ is the integer representing the discrete time evolution. The LE for continuous and discrete-time systems are obtained, respectively,  from
\begin{equation}
    \lambda = \lim_{t\to\infty} \frac{1}{t} \ln{|J(t)|},\qquad\qquad  \lambda =  \lim_{n\to\infty}\frac{1}{n} \ln{\left(\prod_i^n|J_i|\right)}.
\end{equation}
Crucial to mention is that the above LE stability is relative to small perturbation under \textit{translation} of the initial condition. 

In the present work, the $q$-Lyapunov stability is analyzed and obtained from the \textit{dilatation} and \textit{contraction} of initial conditions.
Using the derivate from Eq.~(\ref{df}), it is possible to construct the corresponding linearized equations
$d{\bf y}^{(q)}/dt=J^{(q)}\, {\bf y}^{(q)}$ by using the $q$-Jacobian,
whose eigenvalues provide the stability of trajectories under spatial contraction and dilatation. In other words, the $q$-LE ($\lambda^{(q)}$) for continuous and discrete-time dynamical systems are obtained, respectively, from 
\begin{equation}
    \lambda^{(q)} = \lim_{t\to\infty} \frac{1}{t} \ln{|J^{(q)}(t)|},\qquad\qquad  \lambda^{(q)} =  \lim_{n\to\infty}\frac{1}{n} \ln{\left(\prod_i^n|J_i^{(q)}| \right)}.
\end{equation}

It is necessary to mention that the usual Lyapunov stability and attractors in dissipative systems are invariant under dilatation and contraction of coordinates. However, crucial to realize is that the dilatation and contraction of coordinates considered here is a consequence of the $q$-derivative proposed by \cite{ayse97}, and is therefore not the usual simple linear dilatation or contraction of the type $q x$. Furthermore, in distinction to the usual LE for which quadratic displacements are neglected, in the $q$-derivative, higher-order dilatation and contraction become relevant.

\section{Results}
\label{results}

In this Section, we consider the contraction and dilatation of coordinates in the dissipative chaotic Hénon map in Sec.~\ref{HM}, and in the two nonintegrable conservative systems, the Hénon-Heiles potential in Sec.~\ref{HH} and the diamagnetic problem in Sec.\ref{DD}. The nature of regular and chaotic motion is assessed by calculating the maximal $q$-Lyapunov exponent ($q$-LE); regular motion is indicated by a vanishing value for the maximal $q$-LE, while a positive value signifies chaos. The maximal $q$-LE is computed numerically using the standard method from \cite{wolf85}.

\subsection{The dissipative case - the Hénon map}
\label{HM}

The familiar two-parameter Hénon map is a representative example of phenomena observed in several other maps and is given by \cite{henon76}
\begin{eqnarray}
    x_{n+1} & = & a-x_n^2+b\,y_n,\cr
    & \label{Hmap}\\
    y_{n+1} & = & x_n, \nonumber
    \end{eqnarray}
with ($a,b$) being parameters, ($x_n,y_n$) the variables at discrete times $n=0,1,2,\ldots$. The $q$-Jacobian becomes
\begin{eqnarray}
    J^{(q)}_n = \left(  \begin{matrix}
               -[\psi_x^{(2)}]_q\,x_n & b\cr
            1 & 0\cr
                 \end{matrix}
    \right),
\end{eqnarray}
with $[\psi_x^{(2)}]_q = (q^2-1)/(q-1)$ since the degree of $x_{n}$ in the first equation from (\ref{Hmap}) is $\psi_x=2$. In this model, the dissipation remains unaffected under contraction or dilatation of coordinates, since the determinant $|J_q|=-b$ is independent of $q$. Furthermore, we have two $q$-LEs, $\lambda_1^{(q)}>\lambda_2^{(q)}$, which satisfy $\lambda_1^{(q)}+\lambda_2^{(q)}=-b$.  Note that for $\lim_{q\to1}[\psi_x^{(2)}]_q=2$ the usual Jacobian is recovered.

\begin{figure}[H]
        \centering
        \includegraphics[width = 0.7\textwidth,height=0.3\textheight]{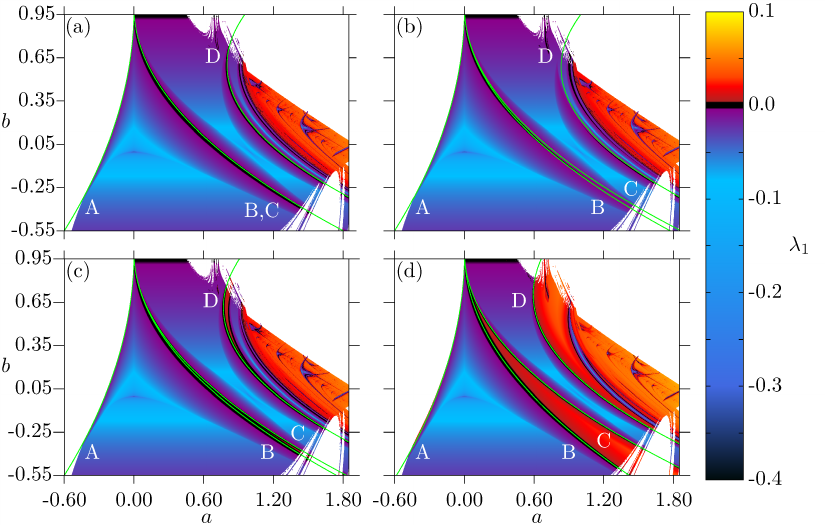}
        \caption{$q$-Lyapunov exponent in the parameter space ($a,b$) for (a)  the usual case $q=1.0$, (b) $q=0.95$, (c) $q=1.05$ and (d) $q=1.4$. {The green curves are the birth of period 1 and the bifurcation curves $P_{12}, P_{1'2'}$ and $P_{24}$, denoted as A, B, C and D, respectively.} }
        \label{fig1}
    \end{figure}

For the first and second iterations of the map (\ref{Hmap}) and $J^{(q)}_n$, analytical curves can be found in the parameter space ($a,b,q$) separating stable periodic motion with periods $1,2$ and $4$. Following the procedure described in \cite{gallas95}, it is straightforward to derive the curves
\begin{eqnarray}
&& P_1=a(q+1)^2+q(1-b)^2=0,\hspace{1.2cm} \mbox{(Birth of period $1$)}\cr
&&\cr
& & P_{12}=a(q+1)^2-(2+q)(1-b)^2  =0, \quad \mbox{(Bifurcation $1\to2$)}\cr
&&\cr
& & P_{1'2'}=a(q+1)^2-q(2+q)(1-b)^2  =0, \quad \mbox{(Bifurcation $1'\to2'$)}\cr
&&\cr
& & P_{24}=a(q+1)^2-q(2+q)(1-b)^2- 2(1+b^2) =0. \quad \mbox{(Bifurcation $2\to4$)} \cr
\nonumber
\end{eqnarray}
 The $P_{12},P_{1'2'}$ and $P_{24}$ curves are the boundaries between stable periodic motions in the parameter space ($a,b,q$) as the period-doubling bifurcation occurs. For high periods, such analytical curves and arbitrary $q$-values are impossible to obtain. In such cases, we rely on numerical simulations. For $q=1$, the above expressions reduce to the curves derived in \cite{zele85,gallas95,beims-gallas97}, where the curves  $P_{12}$ and $P_{1'2'}$ are degenerated.    Figure \ref{fig1}  shows the maximal $q$-LE, $\lambda_1^{(q)}$, as a function of the parameters ($a,b$) and distinct values of $q$. Red to yellow colors for increasing positive $\lambda_1^{(q)}$, and purple to light and dark blue for increasing negative values of the $\lambda_1^{(q)}$. Black color for the bifurcating points with vanishing $\lambda_1^{(q)}$. The green curves are {the birth of period 1 (A)}, curves $P_{12}$ (B), $P_{1'2'}$ (C) and $P_{24}$ (D) and match with the zero $q$-LEs.  The usual case $q=1$ is shown in Fig.~\ref{fig1}(a) and is a well-known result that includes isoperiodic stable structures ("shrimps") embedded in the chaotic regime \cite{gallas93}. The effect of $q\ne 1$ is shown in Figs.~\ref{fig1}(b)-(d). Figure \ref{fig1}(b) displays the $\lambda_1^{(q)}$ for the contraction case with $q=0.6$. It is visible that the degenerated curve $P_{12}$ is splitted in two cases, $P_{12}$ (B) and $P_{1'2'}$ (C). Analogous splitting of the curves occurs for the dilatation $q=1.05$ and $1.4$, shown respectively in Figs.~\ref{fig1}(c) and (d).  

We observe that for the dilatation, the chaotic motion emerges between the curves $P_{12}$ (B) and $P_{1'2'}$ (C), increasing the overall amount of chaotic motion in the parameter space. Besides, the amount of positive $q$-LEs also increases on the right of the curve $P_{24}$ (D), significantly changing the structures embedded in the chaotic regime.
\begin{widetext}
\begin{figure}[H]
        \centering
                 \includegraphics[scale=0.8]{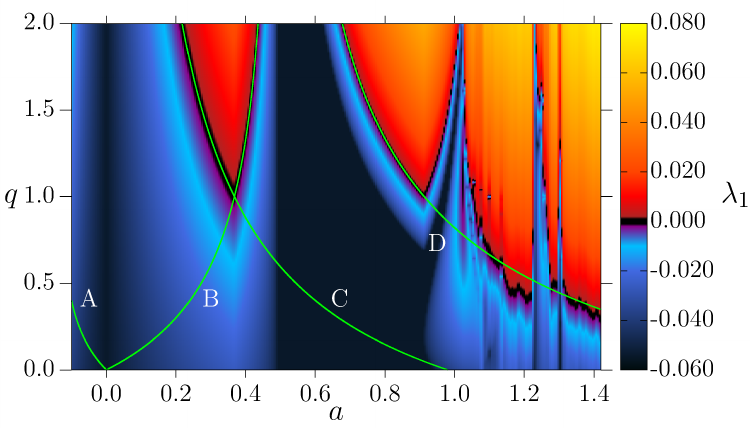}  
        \caption{$q$-Lyapunov exponent in the parameter space ($a,q$) for $b=0.3$. 
        {The green curves are the birth of period 1, and the bifurcation curves 
        $P_{12}, P_{1'2'}$ and $P_{24}$, denoted as A, B, C and D, respectively.} }
        \label{fig2}
    \end{figure}
      \end{widetext}

Figure \ref{fig2} displays the maximal $q$-LE (see color bar) in the parameter space ($a,q$) for $b=0.3$. For $q=1$, we have the usual period-doubling bifurcation diagram as a function of $a$. The curves $P_{12}$ (B) and $P_{1'2'}$ (C) cross at $q=1$ where the usual bifurcation from period $1\to 2$ occurs. This crossing point is the origin of the chaotic motion for $q>1$. Similar behavior is expected for $P_{24}$ (D) at $q=1$, where the usual bifurcation from period $2\to 4$ occurs. However, we could not derive an analytical result for $P_{2'4'}$ to confirm the crossing.  Even though, chaotic motion for $q>1$ occurs when $P_{24}$ (D) crosses $q=1$. Figure \ref{fig2} demonstrates that the regular dynamics for $q=1$ becomes irregular for $q>1$ \textit{at the bifurcation points} from $q=1$. In other words, the bifurcating points from $q=1$ are unstable under the dilatation of the coordinates (see below). On the contrary, the bifurcating points become more stable (smaller $q$-LE) under the contraction of the coordinates. For larger values of $a$ and $q=1$, many trajectories are chaotic, besides some periodic windows. While the dilatation of coordinates ($q>1$) increases the $q$-LEs, the contraction  ($q<1$) decreases the $q$-LEs of the unstable trajectories until they become stable. Furthermore, the contraction of coordinates enlarge the periodic windows.

As mentioned, the bifurcating points, for which the usual LE vanishes, become unstable due to the dilatation ($q>1$). It is known in the linear stability analysis that when the LE vanishes, the first derivative is zero, and second-order terms are needed to describe the local dynamics around the bifurcating points. From our analytical and numerical results, the dilatation is providing such a second-order contribution, as can be observed in Fig.~\ref{fig2} for $q>1$.

\subsection{The conservative cases}

For both nonintegrable conservative problems considered in this Section, the numerical integration uses the fourth-order Runge-Kutta integrator with an adaptable step. In addition to the maximal $q$-LE we also analyse the regular and chaotic motion by constructing the Poincaré Surface of Section (PSS).

\subsubsection{The Hénon-Heiles problem}
\label{HH}

The Hénon-Heiles (H-H) problem is a classic example of a dynamical system that exhibits chaotic behavior. It was introduced by M.~Hénon and C.~Heiles in 1964 \cite{heiles64} as a simplified model to study the motion of stars in a galaxy. The Hamiltonian is given by 
\begin{equation}
    H_{\mbox{\tiny H}}=\frac{1}{2} \left(p_x^2+p_y^2\right)+ \frac{1}{2} \left(Ax^2+By^2\right) +Dx^2y - \frac{C}{3}y^3,
    \label{HHH}
\end{equation}
where $A,B,C$ and $D$ are constants, and $ H_{\mbox{\tiny H}}=E_{\mbox{\tiny H}}$ is the total conserved energy.  The presence of nonlinear and coupling terms in the potential leads to complex and chaotic behavior in the motion of particles governed by the H-H problem. It has been extensively studied in the context of dynamical systems theory and chaos theory and is well known to be integrable for three cases \cite{chang81,chang82,gramma82,saha86}:
{(i) $A=B$ and $C=-D$ 
(ii) $A,B$ arbitrary and $C=-6D$, and (iii) $B=16A$ and $C=-16C$.
For other relations between $A,B,C$, and $D$, the dynamic is mixed with regular and chaotic motions, depending on the values of these constants and the total energy $E_{\mbox{\tiny H}}$.  The possibility of continuously changing the parameters $A,B,C,D$ to obtain integrable (regular), mixed and chaotic dynamics transforms the H-H problem of major interest. Hamilton's equations of motion are given by
\begin{eqnarray}
    \dot x &=& \frac{\partial H_{\mbox{\tiny H}}}{\partial p_x} = p_x, 
    \quad \dot y = \frac{\partial H_{\mbox{\tiny H}}}{\partial p_y}   = p_y,\label{f2}\\
   & & \cr
         \dot p_x &= & -\frac{\partial H_{\mbox{\tiny H}}}{\partial x}= - A\,x-2D\,xy,\label{f3}\\
   & & \cr
    \dot p_y &=&  -\frac{\partial H_{\mbox{\tiny H}}}{\partial y} =- B\,y - D\,x^2 + C\,y^2,\label{f4}\\ \nonumber
\end{eqnarray}
and the associated $q$-Jacobian becomes
\begin{eqnarray}
    J_q= \left(  \begin{matrix}
               0 & 0 &  1\, & 0\cr
               0 & 0 & 0 &  1\cr
             -A-2D\,y & -2D\,x & 0 & 0 \cr
               -D[\psi_x^{(2)}]_q\,x &  -B + C[\psi_y^{(2)}]_q y & 0 & 0  \end{matrix}
    \right),\nonumber
\end{eqnarray}
with $[\psi_x^{(2)}]_q = [\psi_y^{(2)}]_q = (q^2-1)/(q-1)$.

Figure \ref{fig3} displays the PSS for three distinct energies $E_{\mbox{\tiny H}} = 1/8$ (left column), $1/7$ (middle column) and $1/6$ (right column), and three different values of $q=0.5$ (top row), $q=1.0$ (middle row)  and $1.5$ (bottom row). The values of the $q$-LEs are provided by the colorbar. Lilac for vanishing $q$-LEs, and dark to light blue, yellow to red for increasing positive $q$-LEs. For $q=1$, we have the usual case. Islands and tori have a negative $q$-LE, while inside the chaotic sea, the $q$-LEs are positive. First, we observe that for smaller energies, the amount of regular motion increases (as well-known) and the $q$-LEs decrease for regular and chaotic motion.   For the contraction, $q=0.5$, the values of the positive $q$-LEs for the chaotic trajectory decrease. For the dilatation, $q=1.5$, the values of the positive $q$-LEs increase. For trajectories with vanishing $q$-LEs for $q=1.0$, the effect of $q$ is more complicated.
\begin{figure}[H]
        \centering
       \includegraphics[scale=0.8] {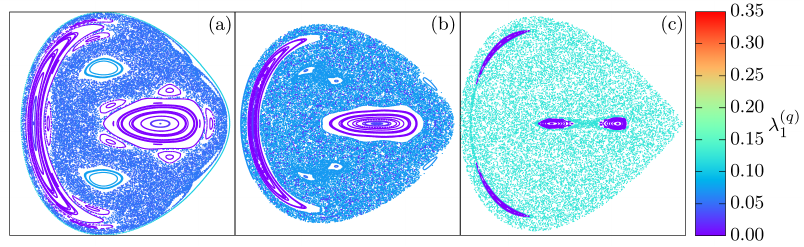}
             \includegraphics[scale=0.8] {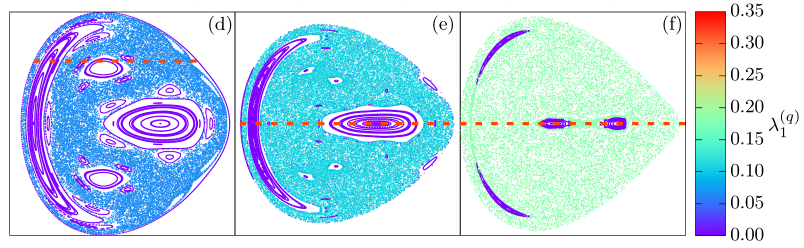}
                   \includegraphics[scale=0.8] {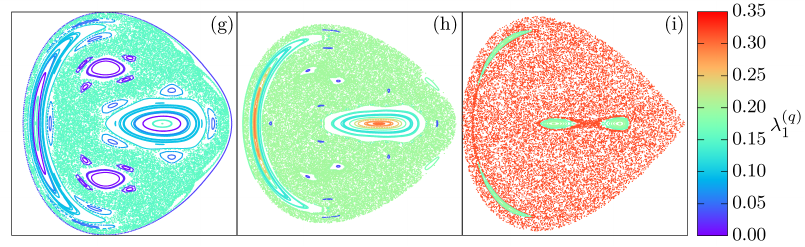}
        \caption{Poincaré surface of section $(y,p_y)$ for three energies, $E_{\mbox{\tiny H}} = 1/8$ (left column), $1/7$ (middle column) and $1/6$ (right column). Top row for $q=0.5$, middle row for $q=1$, and bottom row for $q=0.5$. The color bar shows the value of the $q$-LEs. The intervals are $-0.6 \le p_{y} \le 0.6$ for all three energies and  $-0.5 \le y \le 1.0$, $-0.5 \le y \le 0.8$ and $-0.5 \le y \le 0.7$, for energy values $1/6$, $1/7$ and $1/8$, respectively.}
        \label{fig3}
    \end{figure}

A quantitative way to observe the effect of $q$ on the LEs is to plot them as a function of the ICs. This is shown in Fig.~\ref{fig33} for representative values of $q=1.0$ (black points), $q=1.5$ (red points), $q=0.5$ (blue points) and for three energies  $E_{\mbox{\tiny H}} = 1/8$ (left panel), $1/7$ (middle panel) and $1/6$ (right panel). In this case, the ICs are along the red dashed lines shown in Figs.~\ref{fig3} (d),(e) and (f). The general observed tendency is that for those ICs that lead to positive $q$-LEs for $q=1.0$, dilatation increases the $q$-LEs, while contraction decreases the $q$-LEs. This is more evident in Fig.~\ref{fig33}(c), but is also observed in Figs.~\ref{fig33}(a)-(b) for all ICs which for $q=1$ lead to positive $q$-LEs. On the other hand, for ICs with vanishing $q$-LEs for $q=1$, contraction \textit{and} dilatation tends to increase the $q$-LEs. However, there are some exceptions, for example, in the middle of Fig.~\ref{fig33}(c), where the blue points assume vanishing $q$-LEs after contraction. Another example is the middle of Fig.~\ref{fig33}(a), where the $q$-LEs for dilatation are smaller than those for contraction. In this case, the elliptic point is related to a period-$2$ orbit, and the $q$-LE stability under dilatation and contraction of the surrounding tori are distinct from the other regions in the PSS. Summarizing, the effect of dilatation and contraction of coordinates on chaotic trajectories seems to be simple, but the resulting $q$-LEs are not trivial when the perturbed trajectories are tori.  
\begin{figure}[H]
        \centering
        \includegraphics[scale=1.0] {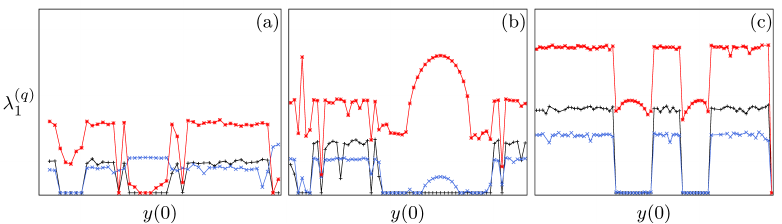}
        \caption{The $q$-LE as a function of the IC ($y(0)$) for three distinct values $q=1$ (black points), $q=0.5$ (blue points), $q=1.5$ (red points) and for three energies, $E_{\mbox{\tiny H}} = 1/8$ (left column), $1/7$ (middle column) and $1/6$ (right column). The intervals are $0.0 \le \lambda_{1}^{(q)} \le 0.33$ for the three energies and  [$-0.45 \le y(0) \le 0.53, p_y(0)=0.28$], [$-0.46 \le y(0) \le 0.48, p_y(0)=0.0$] and [$-0.36 \le y(0) \le 0.48, p_y(0)=0.0$], for energy values $1/6$, $1/7$ and $1/8$, respectively. The ICs are chosen along the long-dashed red lines from Fig.~\ref{fig3}.}
        \label{fig33}
    \end{figure}

Figure \ref{fig4} displays the positive $q$-LE as a function of the energy and for three distinct values of $q$, namely red points for $q=1.5$, black points for $q=1.0$ and blue points for $q=0.5$. Roughly speaking, the $q$-LE increases as a power-law with the energy, whose exponent does not change with $q$. A fit demonstrates the $\lambda= E^{\beta}$ for the three curves, where $\beta\sim 2.99$ for $q=0.5$, $\beta\sim 3.45$ for $q=1.0$ and $\beta\sim 2.38$ for $q=1.5$.
\begin{figure}[H]
        \centering
        \includegraphics[scale=0.64] {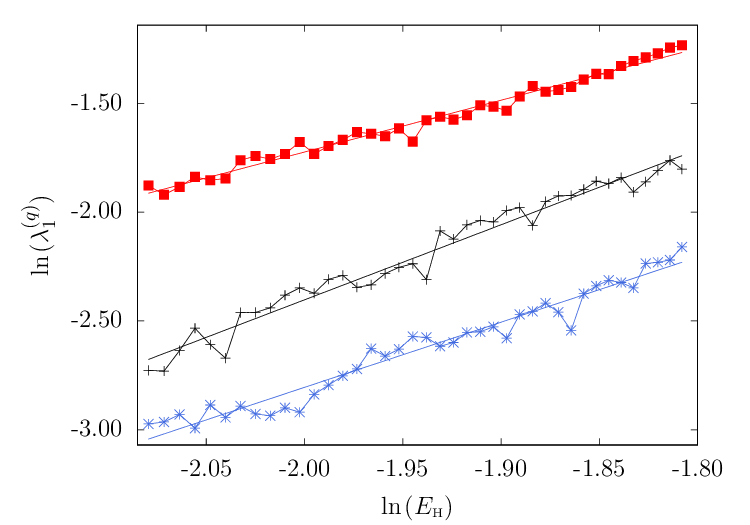}
        \caption{Log-log plot of the $q$-Lyapunov exponent as a function of the energy $E_{\mbox{\tiny H}}$ and distinct values of $q$: red for $q=1.5$, black for $q=1$, and blue for $q=0.5$. Straight lines are the corresponding power-law regressions (see text).}
        \label{fig4}
    \end{figure}
    The linear fit $\ln{(\lambda_1^{(q)})}=c+d \ln{(E_{\mbox{\tiny H}})}$ in the log-log plot is $(c,d) = (3.18142,2.99323)$ for $q=0.5$, $(c,d)= (4.4954,3.4491)$ for $q=1.0$ and $(c,d) = (3.0391,2.3813)$ for $q=1.5$. For the usual case $q=1.0$, the power-law increase of the LEs with the energy agrees with earlier results \cite{shev03}. For $q=1.5$, the power-law is still visible, whose exponent $d$ decreases compared to the usual case. For $q=0.5$, we again use a power-law dependence, but for small energies, we observe some stronger variations of the $q$-LE. The reason for that is the increasing amount of sticky motion for smaller energies. Besides, the local peak observed for $q=0.5$ (blue) at lower energies is a consequence of the sticky motion close to the tori around the period-$2$ orbit observed in the left-column PSSs in Fig.~\ref{fig3}. This was checked numerically by plotting solely the chaotic trajectory in the PSS for energies to the left, at, and to the right of the peak. The amount of sticky motion is larger at the peak, and since $q=0.5$ tends to increase the $q$-LEs from tori around the period-$2$ orbit [see Fig.~\ref{fig33}(a)], the peak is expected.

\subsubsection{The diamagnetic problem}
\label{DD}

The Hamiltonian of the diamagnetic Kepler problem with total energy $E_{\mbox{\tiny D}}$ is given by \cite{dieter89}
\begin{equation}
    H_{\mbox{\tiny D}}=\frac{1}{2} \left(p_x^2+p_y^2+p_z^2\right)+\gamma\left(x^2+y^2\right)-\frac{e}{r},
    \label{HU}
\end{equation}
with $\gamma=\frac{e^2B^2}{8mc^2}$, $r=\sqrt{x^2+y^2+z^2}$, $e$ and $m$ are electron's charge and mas $m$, and the intensity of the static magnetic field in the $z$-direction is $B=4.72\times 10^5 T$. The angular momentum $L_z$ around $z$ is a conservative quantity, and we consider $L_z=0$. The Hamiltonian (\ref{HU}) has been studied in the 80's and 90's of the last century in the context of quantum chaos \cite{holle86,kleppner89,kleppner89-2,dieter89,delande91,delos92,beims93,beims96,beims97}. Generalizations using an external additional electric field and a van der Waals interaction were studied analytically and numerically in the classical and quantum cases \cite{beims00}.  The diamagnetic term proportional to $\gamma$ appears inside the square root since it is derived from the generalized momentum $(\textbf{p}-\frac{e}{c}\textbf{A})$, where $\textbf{A}$ is the vector potential with components $A_x=-By/2, A_y=Bx/2$  and $A_z=0$. The magnetic field is then obtained from $\textbf{A} = \textbf{x}\times \textbf{B}/2$. For $\gamma=0$, we obtain the well-known integrable and separable case. For increasing values of $\gamma$, a mixed (regular plus chaotic) motion is developed. A completely chaotic dynamic is obtained for large values of $\gamma $ and can be explored numerically. For this, the singularity of the Coulomb potential at $r=0$ should be avoided, and regularization is needed. This is implemented using semi-parabolic coordinates
\begin{equation}
    x=uv\cos{\theta},\quad
    y=uv\sin{\theta},\quad
    z=\frac{1}{2}\left(u^2-v^2\right),
\end{equation}
with the corresponding canonical momenta
\begin{eqnarray}
    p_x&=& \frac{up_v\cos{\theta}}{u^2+v^2} + \frac{vp_u\cos{\theta}}{u^2+v^2} -\frac{p_{\phi}\sin{\theta}}{uv},\cr
    &&\cr
    p_y&=& \frac{p_{\phi}\cos{\theta}}{uv}+\frac{up_v\sin{\theta}}{u^2+v^2} + \frac{vp_u\sin{\theta}}{u^2+v^2},\cr
    &&\cr
    p_z&=&\frac{up_u - vp_v}{u^2+v^2}.
\end{eqnarray}
The  regularized Hamiltonian becomes
\begin{equation}
    {\cal H}_{\mbox{\tiny D}}(u,v) = 2(u^2+v^2)[H_{\mbox{\tiny D}}(u,v)-E_{\mbox{\tiny D}}]=2Z, 
\end{equation}
with
\begin{eqnarray}
    {\cal H}_{\mbox{\tiny D}}(u,v) &=& \frac{1}{2}\left(p_u^2+p_v^2+\frac{p^2_{\phi}}{u^2v^2}\right) +\gamma(u^2+v^2)u^2v^2+\cr 
    &&\cr
    & &\frac{\varepsilon}{2}(u^2+v^2).
    \label{Huv}
\end{eqnarray}
Here $\varepsilon = -2E_{\mbox{\tiny D}}$, $p_{\phi}$ is a cyclic coordinate and, therefore, a constant of motion which, for simplicity, we consider to be zero. Equations of motion are given by
\begin{eqnarray}
    \dot u &=& p_u, \qquad
    \dot v = p_v, \cr
    & & \cr
      \dot p_u &=& -\varepsilon u - 2\gamma(2u^3v^2+v^4u),\cr 
      & & \cr
    \dot p_v &=&-\varepsilon v - 2\gamma(2v^3u^2+u^4v).\nonumber
\end{eqnarray}
Now, it is possible to construct the $q$-Jacobian given by 
\begin{widetext}
\begin{eqnarray}
    J_q= \left(  \begin{matrix}
               0 & 0 & 1 & 0\cr
               0 & 0 & 0 & 1\cr
             -\varepsilon -4\gamma [\psi_u^{(3)}]_q u^2v^2 -2\gamma v^4 & -4\gamma [\psi_v^{(2)}]_q u^3 v -2\gamma [\psi_v^{(4)}]_qv^3u& 0 & 0 \cr
               -4\gamma [\psi_v^{(2)}]_q v^3 u -2\gamma [\psi_v^{(4)}]_qu^3 v &  -\varepsilon -4\gamma [\psi_u^{(3)}]_q u^2v^2 -2\gamma u^4 & 0 & 0  \end{matrix}
    \right),\nonumber
\end{eqnarray}
\end{widetext}
  

Figure \ref{fig5} presents PSS for different energy values ($\varepsilon$) and $q$. Specifically, the left column corresponds to $\varepsilon = 3.0$, the middle column to $\varepsilon = 2.0$, and the right column to $\varepsilon = 1.0$. The rows show values of $q$: the top row for $q=0.5$ (contraction), the middle row for $q=1.0$ (standard case), and the bottom row for $q=1.5$ (dilatation). The color bar indicates the values of the $q$-Lyapunov exponents ($q$-LEs) for each PSS. In the middle row, representing the standard case at $q=1$, a mix of regular and chaotic dynamics is evident at $\varepsilon = 3.0$. As the energy decreases, the prevalence of regular dynamics reduces, and for $\varepsilon = 1.0$, the system exhibits fully chaotic behavior. When comparing the $q=1$ case to the contraction scenario ($q=0.5$, top row), we observe that for initially chaotic trajectories at $q=1$, the $q$-LE values decrease, indicating a stabilizing effect under contraction. In contrast, for initially regular trajectories at $q=1$, the $q$-LE values tend to increase, showing that contraction can destabilize regular dynamics. In the dilatation case ($q=1.5$, bottom row), both regular and chaotic trajectories from the $q=1$ scenario exhibit an increase in $q$-LEs, indicating that dilatation generally destabilizes the dynamics regardless of the initial trajectory type. This behavior is consistent with the patterns observed in the Hénon-Heiles system, where dilatation also intensified instability in both chaotic and regular trajectories.

These findings illustrate how changes in $q$ can modulate the system’s stability across different energy levels, aligning with the previously observed behavior in the H-H problem.

\begin{figure}[H]
        \centering
       \includegraphics[scale=1.00] {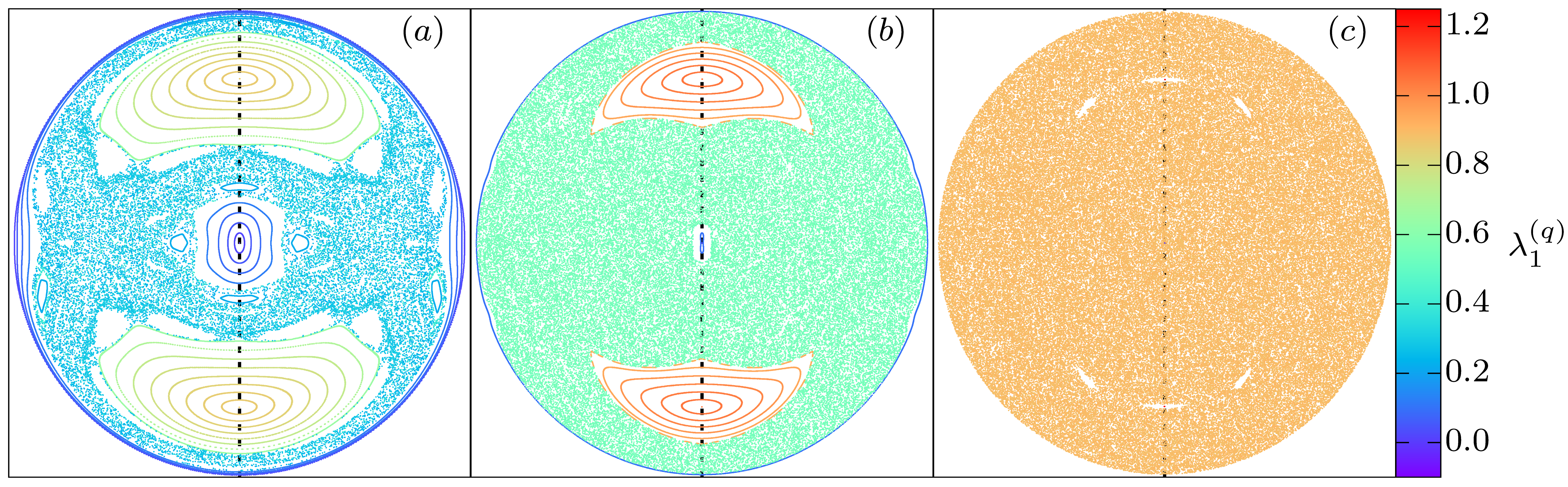}
             \includegraphics[scale=1.00] {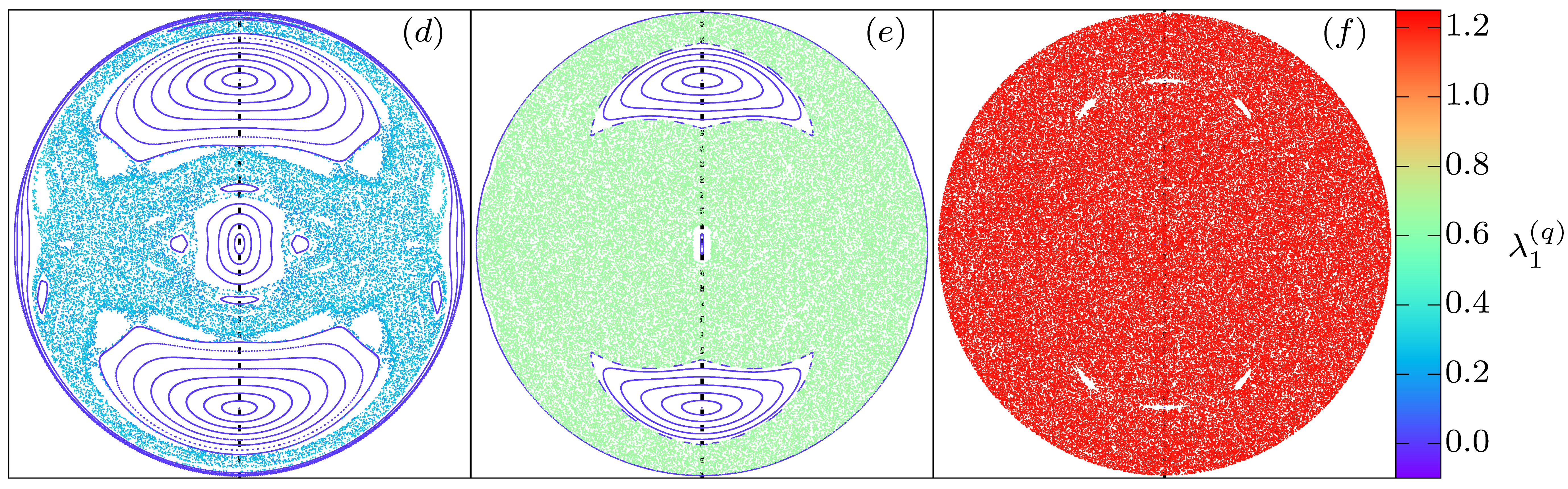}
                   \includegraphics[scale=1.00] {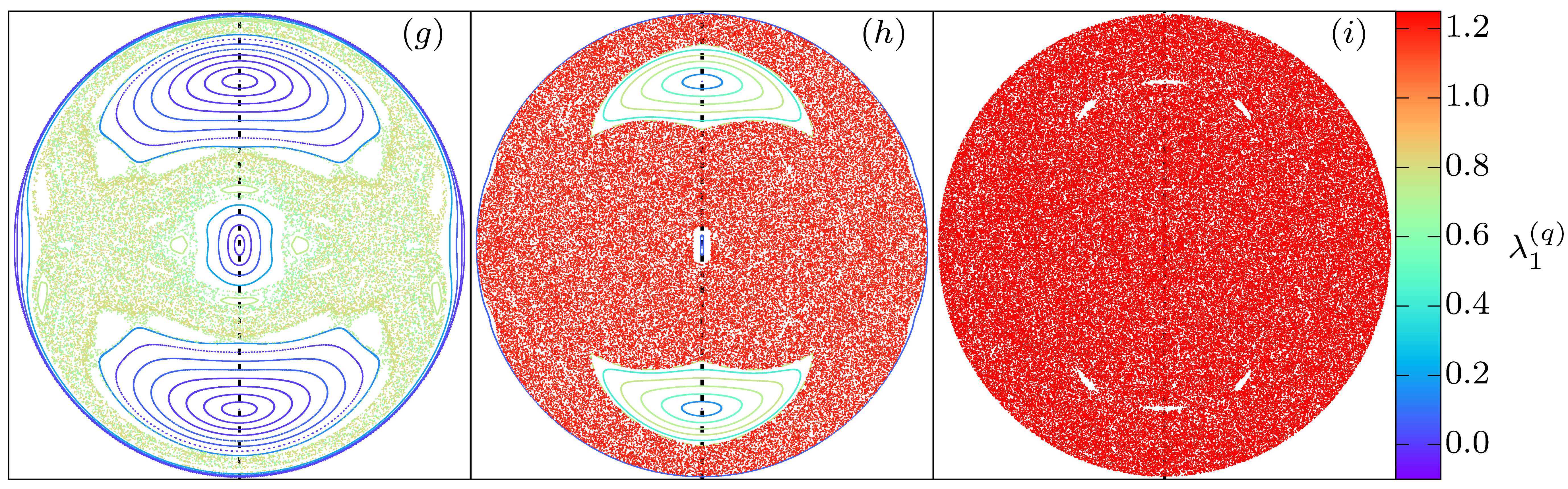}
        \caption{Poincaré surface of section $(v,p_v)$ for three energies, $\varepsilon = 3.0$ (left column), $2.0$ (middle column) and $1.0$ (right column). Top row for $q=0.5$, middle row for $q=1$, and bottom row for $q=1.5$. The color bar shows the value of the $q$-LEs. The intervals are $-2.05 \le p_{v} \le 2.05$ for all three energies and  $-1.20 \le v \le 1.20$, $-1.47 \le v \le 1.47$ and $-2.07 \le v \le 2.07$, for energy values $3$, $2$ and $1$, respectively.}
        \label{fig5}
\end{figure}

Figure \ref{fig6} shows the $q$-LEs for representative values of $q$: black points for $q=1.0$, red points for $q=1.5$ (dilatation), and blue points for $q=0.5$ (contraction). The $q$-LEs are evaluated for three energy levels: $\varepsilon = 3$ (left panel), $\varepsilon = 2$ (middle panel), and $\varepsilon = 1$ (right panel). Here, initial conditions (ICs) are chosen along the vertical dashed lines in Fig.~\ref{fig5}. The general trend observed is that for ICs yielding positive $q$-LEs at $q=1.0$, dilatation (increasing $q$) amplifies the $q$-LEs, while contraction (decreasing $q$) reduces them. This pattern is most prominent in Fig.~\ref{fig6}(c) but is also consistently seen across Figures \ref{fig6}(a) and \ref{fig6}(b) for ICs with positive $q$-LEs at $q=1.0$. In contrast, for ICs with vanishing $q$-LEs at $q=1.0$, both contraction and dilatation tend to increase the $q$-LEs, hinting at a destabilizing effect on initially stable trajectories. Notably, there are exceptions to this behavior. For example, in the middle of Fig.~\ref{fig6}(c), the blue points (contraction) show near-zero $q$-LEs, indicating stability even after contraction. Another exception appears in Figure \ref{fig6}(a), where the $q$-LEs for dilatation are actually lower than those for contraction, contrary to the general trend. 

These findings suggest that while the effects of dilatation and contraction on chaotic trajectories may appear straightforward, the resulting $q$-LEs exhibit complex behavior for perturbed tori. This non-trivial response indicates that the influence of coordinate transformation on the stability of tori is sensitive to initial conditions and the energy level, underscoring the nuanced dynamics that emerge under contraction and dilatation.
   
\begin{figure}[H]
        \centering
        \includegraphics[scale=1.2] {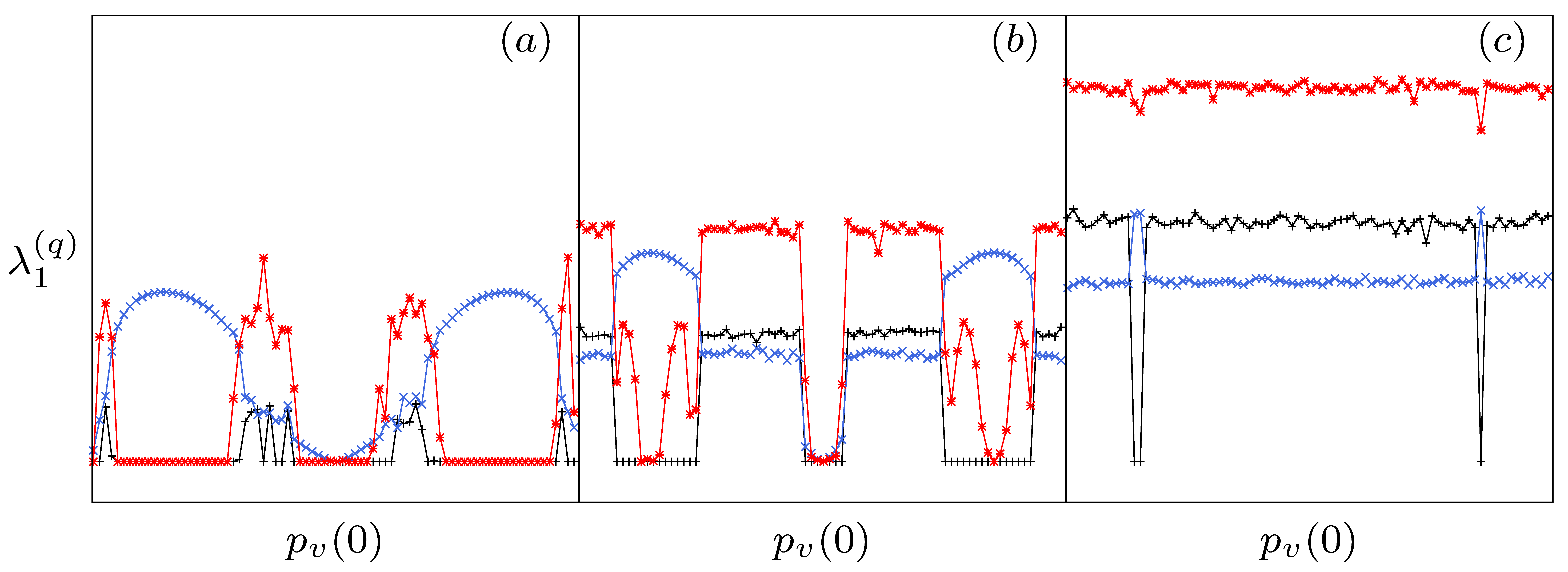}
        \caption{The $q$-LE as a function of the IC ($p_v(0)$ for three distinct values $q=1.0$ (black points), $q=0.5$ (blue points), $q=1.5$ (red points) and for three energies, (a) $\varepsilon = 3.0 $, (b) $\varepsilon =2.0$ and (c) $\varepsilon =1.0$. The intervals are $-0.20 \le \lambda_{1}^{(q)} \le 2.20$ and $-2.00 \le p_v(0) \le 2.00$ for the three energies and $v(0)=0.0$.}
        \label{fig6}
    \end{figure}

In Fig.~\ref{fig7}, we observe the behavior of the positive $q$-LE as a function of energy for three values of $q$: $q=0.5$ (blue points), $q=1.0$ (black points), and $q=1.5$ (red points). Notably, this system demonstrates a linear dependence of the $q$-LE on energy, contrasting with the non-linear behavior seen in the Hénon-Heiles case discussed previously.

        \begin{figure}[H]
        \centering
        \includegraphics[scale=0.8] {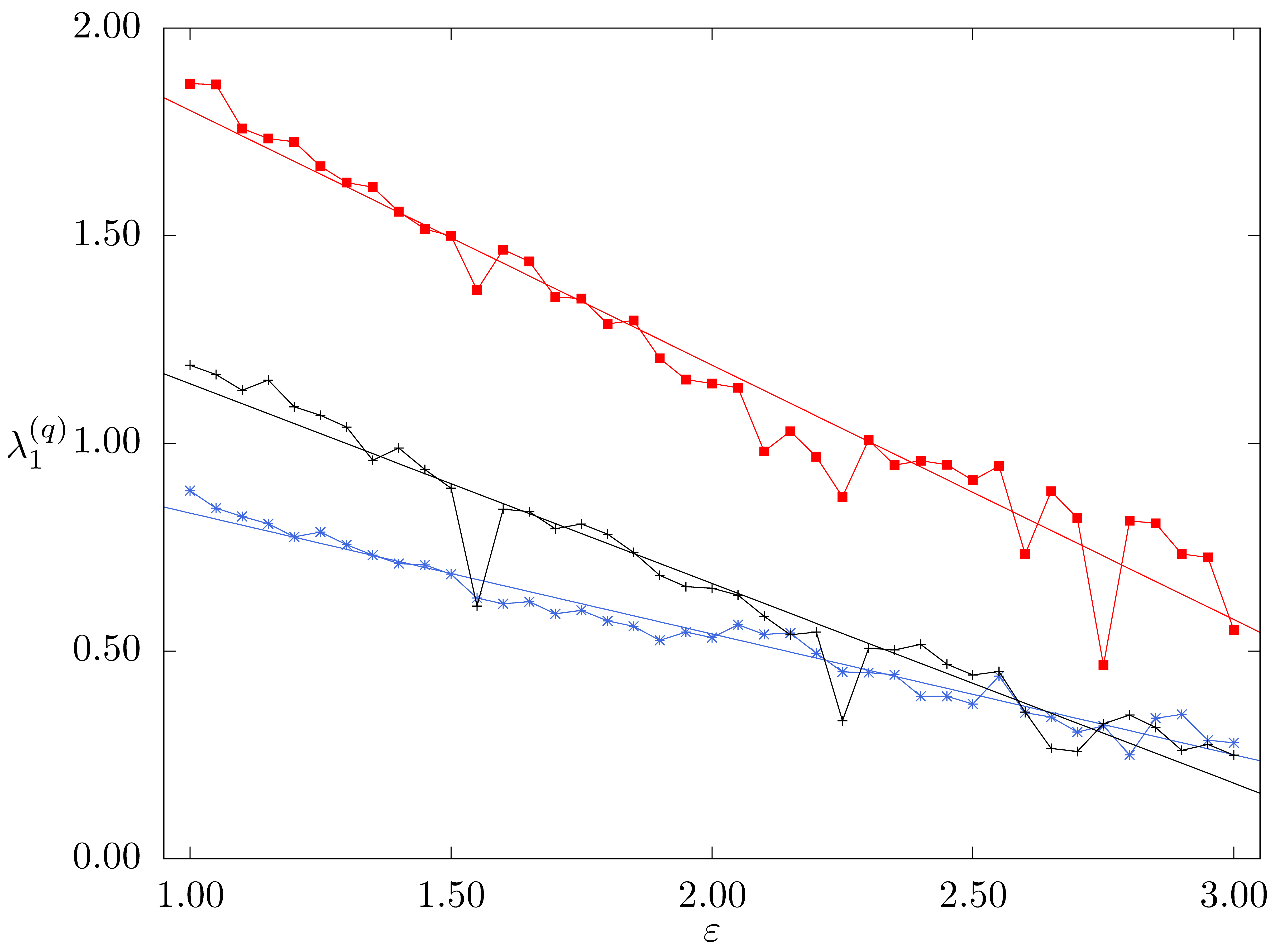}
        \caption{Plot of the $q$-Lyapunov exponent as a function of the energy $\varepsilon$ and distinct values of $q$: red for $q=1.5$, black for $q=1.0$, and blue for $q=0.5$.}
        \label{fig7}
        \end{figure}

The linear relationship of the positive $q$-LE with energy is captured by the fit $\lambda_1^{(q)} = a + b~\varepsilon$, with parameter values $(a, b) = (1.12334, -0.290927)$ for $q=0.5$, $(a, b) = (1.62469, -0.480846)$ for $q=1.0$, and $(a, b) = (2.41462, -0.612911)$ for $q=1.5$. This trend indicates that as $q$ increases, both the intercept $a$ and the slope magnitude $|b|$ increase, suggesting an amplified sensitivity of the $q$-LE to changes in energy.

Additionally, fluctuations in the positive $q$-LE across different energy values reveal rich dynamical features, reflecting a more intricate underlying structure in the system’s response to varying energies. Remarkably, the lines for $q=0.5$ and $q=1.0$ intersect at higher energies, implying that for sufficiently large energy values, the contraction effect associated with $q=0.5$ no longer reduces the positive $q$-LE below that of the $q=1.0$ case. This crossing indicates a limit to the stabilizing influence of contraction at high energies, which could suggest a threshold beyond which the effects of contraction and dilatation converge, or even invert, in terms of their impact on the system's stability.

These findings offer insights into the role of $q$ in modulating stability across energy levels and suggest that the choice of $q$ could potentially be tailored to control stability in various dynamical regimes. Further analysis might explore whether this crossing behavior is a general characteristic of systems with linear energy dependence in their $q$-LE or specific to the system under study.

\section{Conclusion}
\label{conclusion}

In this paper, we demonstrate how the dilatation and contraction of coordinates affect the $q$-Lyapunov stability of trajectories in dynamical systems. To accomplish this, we apply a $q$-deformed derivative to the system’s Jacobian, analyzing three representative models: the dissipative Hénon map, the conservative Hénon-Heiles potential, and the diamagnetic Kepler problem. $q$-Lyapunov stability in each system is evaluated for various values of $q$.

For the dissipative Hénon map, we investigate the impact of $q \neq 1$ within the parameter space ($a, b$), deriving analytical results for low-periodic orbits. These results reveal how the curves for the birth of period-1 points, as well as bifurcations such as period-1 to 2 and period-2 to 4, evolve with $q$. Interestingly, our findings expose a hidden degeneracy in these curves, undetected in previous studies at $q=1$ \cite{zele85,gallas95,beims-gallas97}, providing strong support for the application of dilatation and contraction in $q$-Lyapunov stability analysis. Additionally, numerical simulations in the parameter space illustrate that coordinate dilatation induces chaotic behavior at bifurcation points associated with the $q=1$ case, suggesting that bifurcation points from $q=1$ become destabilized under dilatation.

In conservative systems, we explore the effects of $q \neq 1$ through several Poincaré surfaces of section (PSS). For initial conditions that lead to chaotic trajectories when $q=1$, dilatation increases the Lyapunov exponent (LE), while contraction decreases it, indicating that contractions stabilize, whereas dilatations destabilize chaotic dynamics. In contrast, for initial conditions yielding regular dynamics at $q=1$, both dilatation and contraction raise the $q$-LEs, thereby destabilizing regular motion. Exceptions to these trends are observed, depending on the specific KAM tori for $q=1$, reflecting the classical issue of KAM stability \cite{Lichtenberg}. Particularly irrational tori remain stable under small perturbations, as quantified by the LEs of the Gauss map \cite{beims-gauss}, though we found that the behavior of tori under coordinate transformations cannot be fully explained by the $q$-LEs of the Gauss map.

These findings open the door to potential applications, including links between this framework and relativistic effects involving time dilation and contraction. Future research will explore these connections to further understand the implications of dilatation and contraction in complex dynamical systems.

\acknowledgments
 SZ and VW thank CAPES (Brazilian Agency), TMO, CM and MWB thank the National Council for Scientific and Technological Development—CNPq (Brazilian agency) for financial support (Grant Numbers. 162979/2021-05, 306973/2023-5 and 310294/2022-3, respectively). CM also thanks the Fundação de Amparo à Pesquisa e Inovação do Estado de Santa Catarina—FAPESC (Brazilian agency) for financial support. The authors also acknowledge Prof.~Carlos de Carvalho for computational support at LFTC-DFis-UFPR and LCPAD-UFPR. 


\begin{thebibliography}{37}%
    \makeatletter
    \providecommand \@ifxundefined [1]{%
     \@ifx{#1\undefined}
    }%
    \providecommand \@ifnum [1]{%
     \ifnum #1\expandafter \@firstoftwo
     \else \expandafter \@secondoftwo
     \fi
    }%
    \providecommand \@ifx [1]{%
     \ifx #1\expandafter \@firstoftwo
     \else \expandafter \@secondoftwo
     \fi
    }%
    \providecommand \natexlab [1]{#1}%
    \providecommand \enquote  [1]{``#1''}%
    \providecommand \bibnamefont  [1]{#1}%
    \providecommand \bibfnamefont [1]{#1}%
    \providecommand \citenamefont [1]{#1}%
    \providecommand \href@noop [0]{\@secondoftwo}%
    \providecommand \href [0]{\begingroup \@sanitize@url \@href}%
    \providecommand \@href[1]{\@@startlink{#1}\@@href}%
    \providecommand \@@href[1]{\endgroup#1\@@endlink}%
    \providecommand \@sanitize@url [0]{\catcode `\\12\catcode `\$12\catcode
      `\&12\catcode `\#12\catcode `\^12\catcode `\_12\catcode `\%12\relax}%
    \providecommand \@@startlink[1]{}%
    \providecommand \@@endlink[0]{}%
    \providecommand \url  [0]{\begingroup\@sanitize@url \@url }%
    \providecommand \@url [1]{\endgroup\@href {#1}{\urlprefix }}%
    \providecommand \urlprefix  [0]{URL }%
    \providecommand \Eprint [0]{\href }%
    \providecommand \doibase [0]{http://dx.doi.org/}%
    \providecommand \selectlanguage [0]{\@gobble}%
    \providecommand \bibinfo  [0]{\@secondoftwo}%
    \providecommand \bibfield  [0]{\@secondoftwo}%
    \providecommand \translation [1]{[#1]}%
    \providecommand \BibitemOpen [0]{}%
    \providecommand \bibitemStop [0]{}%
    \providecommand \bibitemNoStop [0]{.\EOS\space}%
    \providecommand \EOS [0]{\spacefactor3000\relax}%
    \providecommand \BibitemShut  [1]{\csname bibitem#1\endcsname}%
    \let\auto@bib@innerbib\@empty
    \bibitem [{\citenamefont {Gauss}(1876)}]{gauss1813}%
      \BibitemOpen
      \bibfield  {author} {\bibinfo {author} {\bibfnamefont {C.~F.}\ \bibnamefont
      {Gauss}},\ }\href@noop {} {\bibfield  {journal} {\bibinfo  {journal}
      {reprinted in Werke}\ }\textbf {\bibinfo {volume} {3}},\ \bibinfo {pages}
      {123} (\bibinfo {year} {1876})}\BibitemShut {NoStop}%
    \bibitem [{\citenamefont {Heine}(1846)}]{heine46}%
      \BibitemOpen
      \bibfield  {author} {\bibinfo {author} {\bibfnamefont {E.}~\bibnamefont
      {Heine}},\ }\href@noop {} {\bibfield  {journal} {\bibinfo  {journal}
      {J.~reine angew.~Math.}\ }\textbf {\bibinfo {volume} {32}},\ \bibinfo {pages}
      {210} (\bibinfo {year} {1846})}\BibitemShut {NoStop}%
    \bibitem [{\citenamefont {Heine}(1847)}]{heine47}%
      \BibitemOpen
      \bibfield  {author} {\bibinfo {author} {\bibfnamefont {E.}~\bibnamefont
      {Heine}},\ }\href@noop {} {\bibfield  {journal} {\bibinfo  {journal}
      {J.~reine angew.~Math.}\ }\textbf {\bibinfo {volume} {34}},\ \bibinfo {pages}
      {285} (\bibinfo {year} {1847})}\BibitemShut {NoStop}%
    \bibitem [{\citenamefont {Heine}(1878)}]{heine78}%
      \BibitemOpen
      \bibfield  {author} {\bibinfo {author} {\bibfnamefont {E.}~\bibnamefont
      {Heine}},\ }\href@noop {} {\emph {\bibinfo {title} {Handbuch der
      Kugelfunktionen, Theorie und Anwendung}}},\ Vol.~\bibinfo {volume} {1}\
      (\bibinfo  {publisher} {Reimer, Berlin},\ \bibinfo {year} {1878})\BibitemShut
      {NoStop}%
    \bibitem [{\citenamefont {Gasper}\ and\ \citenamefont
      {Rahman}(1997)}]{gasper97}%
      \BibitemOpen
      \bibfield  {author} {\bibinfo {author} {\bibfnamefont {G.}~\bibnamefont
      {Gasper}}\ and\ \bibinfo {author} {\bibfnamefont {M.}~\bibnamefont
      {Rahman}},\ }\href@noop {} {\emph {\bibinfo {title} {Hypergeomeytic
      Series}}}\ (\bibinfo  {publisher} {University Press: Cambridge},\ \bibinfo
      {year} {1997})\BibitemShut {NoStop}%
    \bibitem [{\citenamefont {Tsallis}\ \emph {et~al.}(1997)\citenamefont
      {Tsallis}, \citenamefont {Plastino},\ and\ \citenamefont
      {Zheng}}]{tsallis97}%
      \BibitemOpen
      \bibfield  {author} {\bibinfo {author} {\bibfnamefont {C.}~\bibnamefont
      {Tsallis}}, \bibinfo {author} {\bibfnamefont {A.~R.}\ \bibnamefont
      {Plastino}}, \ and\ \bibinfo {author} {\bibfnamefont {W.~M.}\ \bibnamefont
      {Zheng}},\ }\href@noop {} {\bibfield  {journal} {\bibinfo  {journal} {Chaos
      Solitons Fractals}\ }\textbf {\bibinfo {volume} {8}},\ \bibinfo {pages} {885}
      (\bibinfo {year} {1997})}\BibitemShut {NoStop}%
    \bibitem [{\citenamefont {Costa}\ \emph {et~al.}(1997)\citenamefont {Costa},
      \citenamefont {Lyra}, \citenamefont {Plastino},\ and\ \citenamefont
      {Tsallis}}]{lyra97}%
      \BibitemOpen
      \bibfield  {author} {\bibinfo {author} {\bibfnamefont {U.~M.~S.}\
      \bibnamefont {Costa}}, \bibinfo {author} {\bibfnamefont {M.}~\bibnamefont
      {Lyra}}, \bibinfo {author} {\bibfnamefont {A.~R.}\ \bibnamefont {Plastino}},
      \ and\ \bibinfo {author} {\bibfnamefont {C.}~\bibnamefont {Tsallis}},\
      }\href@noop {} {\bibfield  {journal} {\bibinfo  {journal} {Phys.~Rev.~E}\
      }\textbf {\bibinfo {volume} {56}},\ \bibinfo {pages} {245} (\bibinfo {year}
      {1997})}\BibitemShut {NoStop}%
    \bibitem [{\citenamefont {Ayse}(1997)}]{ayse97}%
      \BibitemOpen
      \bibfield  {author} {\bibinfo {author} {\bibfnamefont {E.}~\bibnamefont
      {Ayse}},\ }\href@noop {} {\bibfield  {journal} {\bibinfo  {journal}
      {Phys.~Rev.~Lett.}\ }\textbf {\bibinfo {volume} {78}},\ \bibinfo {pages}
      {3245} (\bibinfo {year} {1997})}\BibitemShut {NoStop}%
    \bibitem [{\citenamefont {Lyra}\ and\ \citenamefont {Tsallis}(1998)}]{lyra98}%
      \BibitemOpen
      \bibfield  {author} {\bibinfo {author} {\bibfnamefont {M.}~\bibnamefont
      {Lyra}}\ and\ \bibinfo {author} {\bibfnamefont {C.}~\bibnamefont {Tsallis}},\
      }\href@noop {} {\bibfield  {journal} {\bibinfo  {journal} {Phys.~Rev.~Lett.}\
      }\textbf {\bibinfo {volume} {80}},\ \bibinfo {pages} {53} (\bibinfo {year}
      {1998})}\BibitemShut {NoStop}%
    \bibitem [{\citenamefont {Jaganathan}\ and\ \citenamefont
      {Sinha}(2005)}]{sinha05}%
      \BibitemOpen
      \bibfield  {author} {\bibinfo {author} {\bibfnamefont {R.}~\bibnamefont
      {Jaganathan}}\ and\ \bibinfo {author} {\bibfnamefont {S.}~\bibnamefont
      {Sinha}},\ }\href@noop {} {\bibfield  {journal} {\bibinfo  {journal}
      {Phys.~Lett.~A}\ }\textbf {\bibinfo {volume} {338}},\ \bibinfo {pages} {277}
      (\bibinfo {year} {2005})}\BibitemShut {NoStop}%
    \bibitem [{\citenamefont {Patidar}\ and\ \citenamefont {Sud}(2009)}]{sud09}%
      \BibitemOpen
      \bibfield  {author} {\bibinfo {author} {\bibfnamefont {V.}~\bibnamefont
      {Patidar}}\ and\ \bibinfo {author} {\bibfnamefont {K.~K.}\ \bibnamefont
      {Sud}},\ }\href@noop {} {\bibfield  {journal} {\bibinfo  {journal} {Nonlinear
      Sci.~Numer.~Simulat.}\ }\textbf {\bibinfo {volume} {14}},\ \bibinfo {pages}
      {827} (\bibinfo {year} {2009})}\BibitemShut {NoStop}%
    \bibitem [{\citenamefont {Rech}\ and\ \citenamefont {Beims}(2010)}]{rech10}%
      \BibitemOpen
      \bibfield  {author} {\bibinfo {author} {\bibfnamefont {P.~C.}\ \bibnamefont
      {Rech}}\ and\ \bibinfo {author} {\bibfnamefont {M.~W.}\ \bibnamefont
      {Beims}},\ }\href@noop {} {\bibfield  {journal} {\bibinfo  {journal}
      {J.~Phys.~: Conf.~Ser.}\ }\textbf {\bibinfo {volume} {246}},\ \bibinfo
      {pages} {012006} (\bibinfo {year} {2010})}\BibitemShut {NoStop}%
    \bibitem [{\citenamefont {Barreira}\ and\ \citenamefont {Pesin}(2022)}]{pesin}%
      \BibitemOpen
      \bibfield  {author} {\bibinfo {author} {\bibfnamefont {L.}~\bibnamefont
      {Barreira}}\ and\ \bibinfo {author} {\bibfnamefont {Y.~B.}\ \bibnamefont
      {Pesin}},\ }\href@noop {} {\emph {\bibinfo {title} {Lyapunov exponents and
      smooth ergodic theory}}}\ (\bibinfo  {publisher} {Providence, Rhode Island},\
      \bibinfo {year} {2022})\BibitemShut {NoStop}%
    \bibitem [{\citenamefont {Wolf}\ \emph {et~al.}(1985)\citenamefont {Wolf},
      \citenamefont {Swift}, \citenamefont {Swinney},\ and\ \citenamefont
      {Vastano}}]{wolf85}%
      \BibitemOpen
      \bibfield  {author} {\bibinfo {author} {\bibfnamefont {A.}~\bibnamefont
      {Wolf}}, \bibinfo {author} {\bibfnamefont {J.~B.}\ \bibnamefont {Swift}},
      \bibinfo {author} {\bibfnamefont {H.~L.}\ \bibnamefont {Swinney}}, \ and\
      \bibinfo {author} {\bibfnamefont {J.~A.}\ \bibnamefont {Vastano}},\
      }\href@noop {} {\bibfield  {journal} {\bibinfo  {journal} {Physica D}\
      }\textbf {\bibinfo {volume} {16}},\ \bibinfo {pages} {285} (\bibinfo {year}
      {1985})}\BibitemShut {NoStop}%
    \bibitem [{\citenamefont {Hénon}(1976)}]{henon76}%
      \BibitemOpen
      \bibfield  {author} {\bibinfo {author} {\bibfnamefont {M.}~\bibnamefont
      {Hénon}},\ }\href@noop {} {\bibfield  {journal} {\bibinfo  {journal}
      {Commun.~Math.~Phys.~}\ }\textbf {\bibinfo {volume} {50}},\ \bibinfo {pages}
      {69} (\bibinfo {year} {1976})}\BibitemShut {NoStop}%
    \bibitem [{\citenamefont {Gallas}(1995)}]{gallas95}%
      \BibitemOpen
      \bibfield  {author} {\bibinfo {author} {\bibfnamefont {J.~A.~C.}\
      \bibnamefont {Gallas}},\ }\href@noop {} {\bibfield  {journal} {\bibinfo
      {journal} {Physica A}\ }\textbf {\bibinfo {volume} {222}},\ \bibinfo {pages}
      {125} (\bibinfo {year} {1995})}\BibitemShut {NoStop}%
    \bibitem [{\citenamefont {Hitzl}\ and\ \citenamefont {Zele}(1985)}]{zele85}%
      \BibitemOpen
      \bibfield  {author} {\bibinfo {author} {\bibfnamefont {D.~L.}\ \bibnamefont
      {Hitzl}}\ and\ \bibinfo {author} {\bibfnamefont {F.}~\bibnamefont {Zele}},\
      }\href@noop {} {\bibfield  {journal} {\bibinfo  {journal} {Physica D}\
      }\textbf {\bibinfo {volume} {14}},\ \bibinfo {pages} {305} (\bibinfo {year}
      {1985})}\BibitemShut {NoStop}%
    \bibitem [{\citenamefont {Beims}\ and\ \citenamefont
      {Gallas}(1997)}]{beims-gallas97}%
      \BibitemOpen
      \bibfield  {author} {\bibinfo {author} {\bibfnamefont {M.~W.}\ \bibnamefont
      {Beims}}\ and\ \bibinfo {author} {\bibfnamefont {J.~A.~C.}\ \bibnamefont
      {Gallas}},\ }\href@noop {} {\bibfield  {journal} {\bibinfo  {journal}
      {Physica A}\ }\textbf {\bibinfo {volume} {238}},\ \bibinfo {pages} {225}
      (\bibinfo {year} {1997})}\BibitemShut {NoStop}%
    \bibitem [{\citenamefont {Gallas}(1993)}]{gallas93}%
      \BibitemOpen
      \bibfield  {author} {\bibinfo {author} {\bibfnamefont {J.~A.~C.}\
      \bibnamefont {Gallas}},\ }\href@noop {} {\bibfield  {journal} {\bibinfo
      {journal} {Phys.~Rev.~Lett.~}\ }\textbf {\bibinfo {volume} {70}},\ \bibinfo
      {pages} {2714} (\bibinfo {year} {1993})}\BibitemShut {NoStop}%
    \bibitem [{\citenamefont {Hénon}\ and\ \citenamefont
      {Heiles}(1964)}]{heiles64}%
      \BibitemOpen
      \bibfield  {author} {\bibinfo {author} {\bibfnamefont {M.}~\bibnamefont
      {Hénon}}\ and\ \bibinfo {author} {\bibfnamefont {C.}~\bibnamefont
      {Heiles}},\ }\href@noop {} {\bibfield  {journal} {\bibinfo  {journal} {The
      Astr.Journal}\ }\textbf {\bibinfo {volume} {69}},\ \bibinfo {pages} {73}
      (\bibinfo {year} {1964})}\BibitemShut {NoStop}%
    \bibitem [{\citenamefont {Chang}\ \emph {et~al.}(1981)\citenamefont {Chang},
      \citenamefont {Tabor}, \citenamefont {Weiss},\ and\ \citenamefont
      {Corliss}}]{chang81}%
      \BibitemOpen
      \bibfield  {author} {\bibinfo {author} {\bibfnamefont {Y.~F.}\ \bibnamefont
      {Chang}}, \bibinfo {author} {\bibfnamefont {M.}~\bibnamefont {Tabor}},
      \bibinfo {author} {\bibfnamefont {J.}~\bibnamefont {Weiss}}, \ and\ \bibinfo
      {author} {\bibfnamefont {C.}~\bibnamefont {Corliss}},\ }\href@noop {}
      {\bibfield  {journal} {\bibinfo  {journal} {Phys.~Lett.~A}\ }\textbf
      {\bibinfo {volume} {85}},\ \bibinfo {pages} {211} (\bibinfo {year}
      {1981})}\BibitemShut {NoStop}%
    \bibitem [{\citenamefont {Chang}\ \emph {et~al.}(1982)\citenamefont {Chang},
      \citenamefont {Tabor},\ and\ \citenamefont {Weiss}}]{chang82}%
      \BibitemOpen
      \bibfield  {author} {\bibinfo {author} {\bibfnamefont {Y.~F.}\ \bibnamefont
      {Chang}}, \bibinfo {author} {\bibfnamefont {M.}~\bibnamefont {Tabor}}, \ and\
      \bibinfo {author} {\bibfnamefont {J.}~\bibnamefont {Weiss}},\ }\href@noop {}
      {\bibfield  {journal} {\bibinfo  {journal} {J.~Math.~Phys.~}\ }\textbf
      {\bibinfo {volume} {23}},\ \bibinfo {pages} {531} (\bibinfo {year}
      {1982})}\BibitemShut {NoStop}%
    \bibitem [{\citenamefont {Grammativos}\ \emph {et~al.}(1982)\citenamefont
      {Grammativos}, \citenamefont {Dorizzi},\ and\ \citenamefont
      {Padjem}}]{gramma82}%
      \BibitemOpen
      \bibfield  {author} {\bibinfo {author} {\bibfnamefont {B.}~\bibnamefont
      {Grammativos}}, \bibinfo {author} {\bibfnamefont {B.}~\bibnamefont
      {Dorizzi}}, \ and\ \bibinfo {author} {\bibfnamefont {R.}~\bibnamefont
      {Padjem}},\ }\href@noop {} {\bibfield  {journal} {\bibinfo  {journal}
      {Phys.~Lett.~}\ }\textbf {\bibinfo {volume} {89A}},\ \bibinfo {pages} {11}
      (\bibinfo {year} {1982})}\BibitemShut {NoStop}%
    \bibitem [{\citenamefont {Sahadevan.}\ and\ \citenamefont
      {Lakshmanan}(1986)}]{saha86}%
      \BibitemOpen
      \bibfield  {author} {\bibinfo {author} {\bibfnamefont {R.}~\bibnamefont
      {Sahadevan.}}\ and\ \bibinfo {author} {\bibfnamefont {M.}~\bibnamefont
      {Lakshmanan}},\ }\href@noop {} {\bibfield  {journal} {\bibinfo  {journal}
      {J.~Phys.~A: Math.~Gen.~}\ }\textbf {\bibinfo {volume} {19}},\ \bibinfo
      {pages} {L949} (\bibinfo {year} {1986})}\BibitemShut {NoStop}%
    \bibitem [{\citenamefont {Shevchenko}\ and\ \citenamefont
      {Mel'nikov}(2003)}]{shev03}%
      \BibitemOpen
      \bibfield  {author} {\bibinfo {author} {\bibfnamefont {I.~I.}\ \bibnamefont
      {Shevchenko}}\ and\ \bibinfo {author} {\bibfnamefont {A.~V.}\ \bibnamefont
      {Mel'nikov}},\ }\href@noop {} {\bibfield  {journal} {\bibinfo  {journal}
      {JETP Letters}\ }\textbf {\bibinfo {volume} {77}},\ \bibinfo {pages} {642}
      (\bibinfo {year} {2003})}\BibitemShut {NoStop}%
    \bibitem [{\citenamefont {Friedrich}\ and\ \citenamefont
      {Wintgen}(1989)}]{dieter89}%
      \BibitemOpen
      \bibfield  {author} {\bibinfo {author} {\bibfnamefont {H.}~\bibnamefont
      {Friedrich}}\ and\ \bibinfo {author} {\bibfnamefont {D.}~\bibnamefont
      {Wintgen}},\ }\href@noop {} {\bibfield  {journal} {\bibinfo  {journal}
      {Phys.~Rep.}\ }\textbf {\bibinfo {volume} {183}},\ \bibinfo {pages} {37}
      (\bibinfo {year} {1989})}\BibitemShut {NoStop}%
    \bibitem [{\citenamefont {Holle}\ \emph {et~al.}(1986)\citenamefont {Holle},
      \citenamefont {Wiebusch}, \citenamefont {Main}, \citenamefont {Hager},
      \citenamefont {Rottke},\ and\ \citenamefont {Welge}}]{holle86}%
      \BibitemOpen
      \bibfield  {author} {\bibinfo {author} {\bibfnamefont {A.}~\bibnamefont
      {Holle}}, \bibinfo {author} {\bibfnamefont {G.}~\bibnamefont {Wiebusch}},
      \bibinfo {author} {\bibfnamefont {J.}~\bibnamefont {Main}}, \bibinfo {author}
      {\bibfnamefont {B.}~\bibnamefont {Hager}}, \bibinfo {author} {\bibfnamefont
      {H.}~\bibnamefont {Rottke}}, \ and\ \bibinfo {author} {\bibfnamefont
      {K.}~\bibnamefont {Welge}},\ }\href@noop {} {\bibfield  {journal} {\bibinfo
      {journal} {Phys.~Rev.~Lett.}\ }\textbf {\bibinfo {volume} {56}},\ \bibinfo
      {pages} {2594} (\bibinfo {year} {1986})}\BibitemShut {NoStop}%
    \bibitem [{\citenamefont {Welch}\ \emph {et~al.}(1989)\citenamefont {Welch},
      \citenamefont {Kash}, \citenamefont {Iu}, \citenamefont {Hsu},\ and\
      \citenamefont {Kleppner}}]{kleppner89}%
      \BibitemOpen
      \bibfield  {author} {\bibinfo {author} {\bibfnamefont {G.}~\bibnamefont
      {Welch}}, \bibinfo {author} {\bibfnamefont {M.}~\bibnamefont {Kash}},
      \bibinfo {author} {\bibfnamefont {C.}~\bibnamefont {Iu}}, \bibinfo {author}
      {\bibfnamefont {L.}~\bibnamefont {Hsu}}, \ and\ \bibinfo {author}
      {\bibfnamefont {D.}~\bibnamefont {Kleppner}},\ }\href@noop {} {\bibfield
      {journal} {\bibinfo  {journal} {Phys.~Rev.~Lett.}\ }\textbf {\bibinfo
      {volume} {62}},\ \bibinfo {pages} {893} (\bibinfo {year} {1989})}\BibitemShut
      {NoStop}%
    \bibitem [{\citenamefont {Kash}\ \emph {et~al.}(1989)\citenamefont {Kash},
      \citenamefont {Hsu},\ and\ \citenamefont {Kleppner}}]{kleppner89-2}%
      \BibitemOpen
      \bibfield  {author} {\bibinfo {author} {\bibfnamefont {M.}~\bibnamefont
      {Kash}}, \bibinfo {author} {\bibfnamefont {L.}~\bibnamefont {Hsu}}, \ and\
      \bibinfo {author} {\bibfnamefont {D.}~\bibnamefont {Kleppner}},\ }\href@noop
      {} {\bibfield  {journal} {\bibinfo  {journal} {Phys.~Rev.~Lett.}\ }\textbf
      {\bibinfo {volume} {63}},\ \bibinfo {pages} {1133} (\bibinfo {year}
      {1989})}\BibitemShut {NoStop}%
    \bibitem [{\citenamefont {Delande}\ \emph {et~al.}(1991)\citenamefont
      {Delande}, \citenamefont {Bommier},\ and\ \citenamefont {Gay}}]{delande91}%
      \BibitemOpen
      \bibfield  {author} {\bibinfo {author} {\bibfnamefont {D.}~\bibnamefont
      {Delande}}, \bibinfo {author} {\bibfnamefont {A.}~\bibnamefont {Bommier}}, \
      and\ \bibinfo {author} {\bibfnamefont {J.}~\bibnamefont {Gay}},\ }\href@noop
      {} {\bibfield  {journal} {\bibinfo  {journal} {Phys.~Rev.~Lett.}\ }\textbf
      {\bibinfo {volume} {66}},\ \bibinfo {pages} {141} (\bibinfo {year}
      {1991})}\BibitemShut {NoStop}%
    \bibitem [{\citenamefont {Mao}\ and\ \citenamefont {Delos}(1992)}]{delos92}%
      \BibitemOpen
      \bibfield  {author} {\bibinfo {author} {\bibfnamefont {J.~M.}\ \bibnamefont
      {Mao}}\ and\ \bibinfo {author} {\bibfnamefont {J.~B.}\ \bibnamefont
      {Delos}},\ }\href@noop {} {\bibfield  {journal} {\bibinfo  {journal}
      {Phys.~Rev.~A}\ }\textbf {\bibinfo {volume} {45}},\ \bibinfo {pages} {1746}
      (\bibinfo {year} {1992})}\BibitemShut {NoStop}%
    \bibitem [{\citenamefont {Beims}\ and\ \citenamefont {Alber}(1993)}]{beims93}%
      \BibitemOpen
      \bibfield  {author} {\bibinfo {author} {\bibfnamefont {M.~W.}\ \bibnamefont
      {Beims}}\ and\ \bibinfo {author} {\bibfnamefont {G.}~\bibnamefont {Alber}},\
      }\href@noop {} {\bibfield  {journal} {\bibinfo  {journal} {Phys.~Rev.~A}\
      }\textbf {\bibinfo {volume} {48}},\ \bibinfo {pages} {3123} (\bibinfo {year}
      {1993})}\BibitemShut {NoStop}%
    \bibitem [{\citenamefont {Beims}\ and\ \citenamefont {Alber}(1996)}]{beims96}%
      \BibitemOpen
      \bibfield  {author} {\bibinfo {author} {\bibfnamefont {M.~W.}\ \bibnamefont
      {Beims}}\ and\ \bibinfo {author} {\bibfnamefont {G.}~\bibnamefont {Alber}},\
      }\href@noop {} {\bibfield  {journal} {\bibinfo  {journal} {J.~Phys.~B}\
      }\textbf {\bibinfo {volume} {29}},\ \bibinfo {pages} {4139} (\bibinfo {year}
      {1996})}\BibitemShut {NoStop}%
    \bibitem [{\citenamefont {Beims}(1997)}]{beims97}%
      \BibitemOpen
      \bibfield  {author} {\bibinfo {author} {\bibfnamefont {M.~W.}\ \bibnamefont
      {Beims}},\ }\href@noop {} {\bibfield  {journal} {\bibinfo  {journal}
      {Phys.~Rev.~A}\ }\textbf {\bibinfo {volume} {56}},\ \bibinfo {pages} {R2503}
      (\bibinfo {year} {1997})}\BibitemShut {NoStop}%
    \bibitem [{\citenamefont {Beims}\ and\ \citenamefont {Gallas}(2000)}]{beims00}%
      \BibitemOpen
      \bibfield  {author} {\bibinfo {author} {\bibfnamefont {M.~W.}\ \bibnamefont
      {Beims}}\ and\ \bibinfo {author} {\bibfnamefont {J.~A.~C.}\ \bibnamefont
      {Gallas}},\ }\href@noop {} {\bibfield  {journal} {\bibinfo  {journal}
      {Phys.~Rev.~A}\ }\textbf {\bibinfo {volume} {62}},\ \bibinfo {pages} {043410}
      (\bibinfo {year} {2000})}\BibitemShut {NoStop}%
    \bibitem [{\citenamefont {Lichtenberg}\ and\ \citenamefont
      {Lieberman}(1992)}]{Lichtenberg}%
      \BibitemOpen
      \bibfield  {author} {\bibinfo {author} {\bibfnamefont {A.~J.}\ \bibnamefont
      {Lichtenberg}}\ and\ \bibinfo {author} {\bibfnamefont {M.~A.}\ \bibnamefont
      {Lieberman}},\ }\href@noop {} {\emph {\bibinfo {title} {Regular and Chaotic
      Dynamics}}}\ (\bibinfo  {publisher} {Springer-Verlag},\ \bibinfo {address}
      {New York},\ \bibinfo {year} {1992})\BibitemShut {NoStop}%
    \bibitem [{\citenamefont {Manchein}\ and\ \citenamefont
      {Beims}(2009)}]{beims-gauss}%
      \BibitemOpen
      \bibfield  {author} {\bibinfo {author} {\bibfnamefont {C.}~\bibnamefont
      {Manchein}}\ and\ \bibinfo {author} {\bibfnamefont {M.~W.}\ \bibnamefont
      {Beims}},\ }\href@noop {} {\bibfield  {journal} {\bibinfo  {journal} {Caos,
      Solitons and Fractals}\ }\textbf {\bibinfo {volume} {39}},\ \bibinfo {pages}
      {2041} (\bibinfo {year} {2009})}\BibitemShut {NoStop}%
    \end{thebibliography}
%

\end{document}